\documentclass[12pt]{article}
\usepackage[utf8]{inputenc}
\usepackage[english]{babel}
\usepackage{amssymb,graphicx}
\pdfoutput=1
\usepackage{graphics}
\usepackage{amsmath,amsfonts}
\usepackage{fullpage}
\usepackage{wrapfig} 
\usepackage{braket}
\usepackage[colorinlistoftodos]{todonotes}

\newcommand{\arcosh}{\mathop{\rm arcosh}\nolimits}

\renewcommand{\tanh}{\mathop{\rm th}\nolimits}

\graphicspath{{images/}{figures/}}

\title{Black bounces and remnants in dilaton gravity}
\author{Maxim Fitkevich$^{1,2}$ \vspace{.2cm}\\
\normalsize\it $^1$ Institute for Nuclear Research of the Russian Academy
of Sciences, \\  
\normalsize \it  60th October Anniversary Prospect, 7a, 117312  Moscow, Russia\\
\normalsize\it $^2$ Moscow Institute of Physics and Technology, \\  
\normalsize \it Institutskiy per., 9, 141701 Dolgoprudny, Moscow Region, Russia\\  
\normalsize \it fitkevich@phystech.edu
}

\begin{document}

\maketitle

\begin{flushright}
INR-TH-2022-004
\end{flushright}
\vspace{-1cm}

\abstract{We propose a family of dilaton gravity models possessing bouncing solutions with interiors connecting separate asymptotically flat regions. We demonstrate that inner Cauchy horizons are stable given certain initial conditions. We study causal structure and evaluate thermodynamic properties of black bounces using Euclidean methods. Extremal bounces have zero temperature and can be considered as remnants. We speculate that quantum fluctuations can dissolve event horizons in case of black bounces providing a possible resolution to the information paradox. } 

%%%%%%%%%%%%%%%%%%%%%%%%%%%%%%%%
\section{Introduction}
\label{sec:intro}
Existence of black holes poses one of the most die-hard riddles in theoretical physics. According to quantum field theory black holes evaporate into thermal radiation \cite{Hawking:1975vcx}. This semiclassical result conflicts with the postulate of unitary quantum evolution \cite{Hawking:1976ra}.

The Gauge/String duality make us believe this contrariety is apparent and black hole evaporation is nothing but a sophisticated scattering process \cite{Harlow:2014yka}. Nevertheless, a conclusive proof is still absent. On the contrary, the AMPS-firewall argument sharpen the information paradox as a no-go theorem reading one of the commonly-believed statements, namely purity of Hawking radiation, validity of the local quantum field theory beyond horizon scale, or ``non-dramatic'' horizon, should be omitted \cite{Almheiri:2012rt}.

This claim is not resistant to possible loopholes. There was recently a disclosure of so-called ``islands'' by using replica wormholes to evaluate unitary form of the Page curve during black hole evaporation \cite{Penington:2019npb}. This achievement stimulated a revival of interest in low-dimensional models of gravity which are perfect playgrounds to verify these new ideas \cite{Almheiri:2019qdq}.

Two-dimensional gravity is considered to be renormalizable suggesting that a solution to the information loss problem is analytically tractable at least in principle. It also allows us to leave aside unimportant higher-dimensional complications. Holography provides a connection to condensed matter models like the SYK spin chain \cite{Trunin:2020vwy}, so it can be potentially verified even in the lab experiments \cite{Pikulin:2017mhj}.

In this paper we investigate dilaton gravity models which are modifications of the prolific CGHS model~\cite{Callan:1992rs},
\begin{equation}\label{eq:cghs-action-n}
S_{\mathrm{CGHS}}=\int d^2x\sqrt{-g}\left(e^{-2\phi}R+4e^{-2\phi}\left((\nabla \phi)^2+\lambda^2\right)-\frac12\sum_{i=1}^N(\nabla f_i)^2\right)\;,
\end{equation}
describing interactions of the metric $g_{\mu\nu}$, dilaton $\phi$, and $N$ massless scalar fields $f_i$. The classical CGHS model is exactly solvable but it is not singularity-free, and adding one-loop corrections from the scalar fields does not resolve this issue \cite{Russo:1992ht,Russo:1992ax}.

It was conjectured that quantum gravity effects should resolve singularity. Instead of considering complicated full dynamics at large curvatures one may apply a ``phenomenological'' limiting condition ${R_{\mu\nu\rho\sigma}}^2<\Lambda^2$ with some dimensionfull Planck scale parameter $\Lambda$ \cite{Markov:1984ii}. Next, one can modify the gravity action~\eqref{eq:cghs-action-n} so that singularity is dynamically avoided. By implementing the limiting curvature condition as a constraint the models were obtained where gravitational collapse stops with the de Sitter core's formation \cite{Frolov:1988vj,Frolov:2021kcv}.

A question remains if such solutions actually appear from the UV-completed theory of quantum gravity. Two-dimensional dilaton models can arise as reductions of higher dimensional gravity $M_2\times K_d\,\mapsto\,M_2'$ upon compactification on some subspace $K_d$. In the string theoretic framework it becomes interesting if such a model represents an effective field theory consistent with underlying quantum gravity or belongs to the so-called swampland \cite{Ooguri:2006in}. Generally, one expects that a two-dimensional model does not necessarily have a consistent higher dimensional counterpart. In this sense two-dimensional theories can have some form of effective or approximate holographic description which is yet to be understood \cite{Narayan:2020pyj}. 

Observations coming from Lorentzian path integral approach insist that regular black holes are incompatible with allowed gravitational effective actions \cite{Giacchini:2021pmr,Knorr:2022kqp}. Therefore, the considered non-singular models could only mimic some essential features of black hole-like objects (e.g. fuzzballs or Planck stars) in the true quantum gravity.

In this paper we modify the gravity action \eqref{eq:cghs-action-n} so that a bounce happens instead of the de Sitter core. Therefore, a given class of models contains vacuum solutions similar to the Bardeen black holes. Classical black bounces are described by non-singular metric but possess event horizons and oscillate in global time connecting infinite number of asymptotically flat patches, cf. Ref.~\cite{Simpson:2018tsi}.

In the considered dilaton gravity models there is a threshold mass $M_{\mathrm{ext}}$ corresponding to the extremal bounces. Non-extremal bounces have masses $M>M_{\mathrm{ext}}$ and solutions with $M<M_{\mathrm{ext}}$ are horizonless. Black bounces are more or less stable against matter perturbations but it can lead to the mass inflation phenomenon \cite{Poisson:1989zz} at the certain choice of initial condition.

We evaluate the temperature and the entropy of black bounces using Euclidean methods. The extremal bounces have zero temperature and do not evaporate, and can be considered as remnants. Mean field approximation reads the remnants are stable but it seems likely they undergo quantum decay. An infinitesimal amount of energy is sufficient to dissolve the event horizon completely and all the matter fallen into a would-be interior returns back after the end of evaporation.

The paper is organized as follows. In Sec.~\ref{sec:ldv-models} we find static classical solutions in the proposed class of models. Sec.~\ref{sec:matter-consid} addresses inclusion of the matter fields. And in Sec.~\ref{sec:discussion} we discuss gathered results and illuminate some prospects.

%%%%%%%%%%%%%%%%%%%%%%%%%%%%%%%%
\section{Deformations of CGHS model}\label{sec:ldv-models}
\subsection{General LDV model}
We describe the class of dilaton gravity models possessing the linear dilaton vacuum (LDV),
\begin{equation}
\phi=-\lambda r\;, \qquad R=0\;, \qquad ds^2=-dt^2+dr^2\;,
\end{equation}
as a solution of field equations. One finds it is provided by the action,
\begin{equation}\label{eq:gen-lin-dil}
S_{\mathrm{LDV}}=\int d^2x\sqrt{-g}\left(W(\phi)R+W''(\phi)\left((\nabla\phi)^2+\lambda^2\right)\right)+S^\mathrm{m}\;,
\end{equation}
where primes denote derivatives of $W(\phi)$ with respect to its argument and $S^\mathrm{m}$ is an action for some matter which we ignore in this Section. The choices $W(\phi)=e^{-2\phi}$ and $W(\phi)=e^{-2\phi}-N\phi/48\pi$ correspond to the CGHS and RST \cite{Russo:1992ax} models respectively.

By varying Eq.~\eqref{eq:gen-lin-dil} with respect to $\phi$ and $g^{\mu\nu}$ one derives field equations,
\begin{gather}
W'(\phi)R=2W''(\phi)\Box\phi+W'''(\phi)\left((\nabla\phi)^2-\lambda^2\right)\;, \label{eq:ldv1}\\
g_{\mu\nu}\left(W''(\phi)((\nabla\phi)^2-\lambda^2)+2W'(\phi)\Box\phi\right)-2W'(\phi)\nabla_\mu\nabla_\nu\phi={T^\mathrm{m}}_{\mu\nu}\;, \label{eq:ldv2}
\end{gather}
where ${T^\mathrm{m}}_{\mu\nu}=(-2/\sqrt{-g})\delta S^\mathrm{m}/\delta g^{\mu\nu}$ is the matter stress tensor. Eqs.~\eqref{eq:ldv1}, \eqref{eq:ldv2} have the one-parametric set of vacuum solutions,
\begin{equation}\label{eq:general-et-bh}
ds^2=-f(r)dt^2+\frac{dr^2}{f(r)}\;, \qquad \phi=-\lambda r\;, \qquad f(r)=1+\frac{M}{\lambda W'(\phi)}\;,
\end{equation}
where $M$ is an integration constant corresponding to the ADM-mass for asymptotically flat spacetimes. Global properties of the solution \eqref{eq:general-et-bh} are determined by a form of the function $W'(\phi)$.

If $f(r)>0$ for all $r$ it defines a spacetime without event horizons with configuration resembling a kink from field theory. If $f(r)$ changes its sign the equation $f(r_{\mathrm{h}})=0$ determines a position of the event horizons.

The Ricci curvature is given by formula $R=-\partial_r^2f(r)$ for the Schwarzschild ansatz~\eqref{eq:general-et-bh}. Therefore, a singularity occurs if $W'(\phi_{\mathrm{s}})=0$ and $W''(\phi_{\mathrm{s}})\neq 0$ at $r=-\phi_{\mathrm{s}}/\lambda$. This observation allows us to construct models~\eqref{eq:gen-lin-dil} possessing non-singular black holes with event horizons by explicit choice of $W(\phi)$. In the next Section we provide a concrete example.

Let us calculate the thermodynamic properties of black holes in the models~\eqref{eq:gen-lin-dil}. The solution contributing to the partition function is a Euclidean continuation of Eq.~\eqref{eq:general-et-bh},
\begin{equation}
{ds_E}^2=f(r){dt_E}^2+\frac{dr^2}{f(r)}\;, \qquad\qquad 0\leq t_E <\beta_H\;,
\end{equation}
which is periodic in imaginary time $t_E=it$ and has no conifold singularity on the event horizon $r=r_{\mathrm{h}}$. The last condition fixes the imaginary time period which is the inverse Hawking temperature $\beta_H={T_H}^{-1}=4\pi/f'(r_{\mathrm{h}})$. Relating the event horizon position with the black hole mass by $M=-\lambda W'(\phi_{\mathrm{h}})$ one finds the temperature
\begin{equation}\label{gen-temp}
T_H=\frac{\lambda^2W''(\phi_{\mathrm{h}})}{4\pi M}\;.
\end{equation}
It appears that the extremal black hole has zero temperature because $W''(\phi_{\mathrm{h}})=0$ in this case. This is a signature of a possible remnant at the end of black hole evaporation.

The black hole entropy is given by
\begin{equation}\label{gen-ent}
S_\mathrm{BH}(M)=\int^M_{M_{\mathrm{ext}}}\frac{dM}{T_H(M)}=4\pi W(\phi_{\mathrm{h}})-4\pi W(\phi_{\mathrm{h,\,ext}})\;,
\end{equation}
where we had taken into account that the lightest black hole has vanishing entropy by fixing limits of integration.

\subsection{Example: sinh-CGHS model}\label{sec:reg-bh}
Let us fix $W(\phi)=-2\sinh(2\phi)$ in Eq.~\eqref{eq:gen-lin-dil}, namely consider the action
\begin{equation}\label{eq:grav-action}
S_{\mathrm{sinh}}=-2\int d^2x\,\sqrt{-g}\sinh(2\phi)\left(R+4(\nabla\phi)^2+4\lambda^2\right)\;,
\end{equation}
which describes two copies of the CGHS model plus corrections at large $|\phi|=\lambda|r|$. Left ($\phi>0$) and right ($\phi<0$) copies interpolate smoothly near the ``core'' region of proper size/duration $\simeq\lambda^{-1}$ across the line $\phi=0$.

\begin{figure}[t]
\centerline{
\hspace{0.2cm}\includegraphics[width=7.36cm]{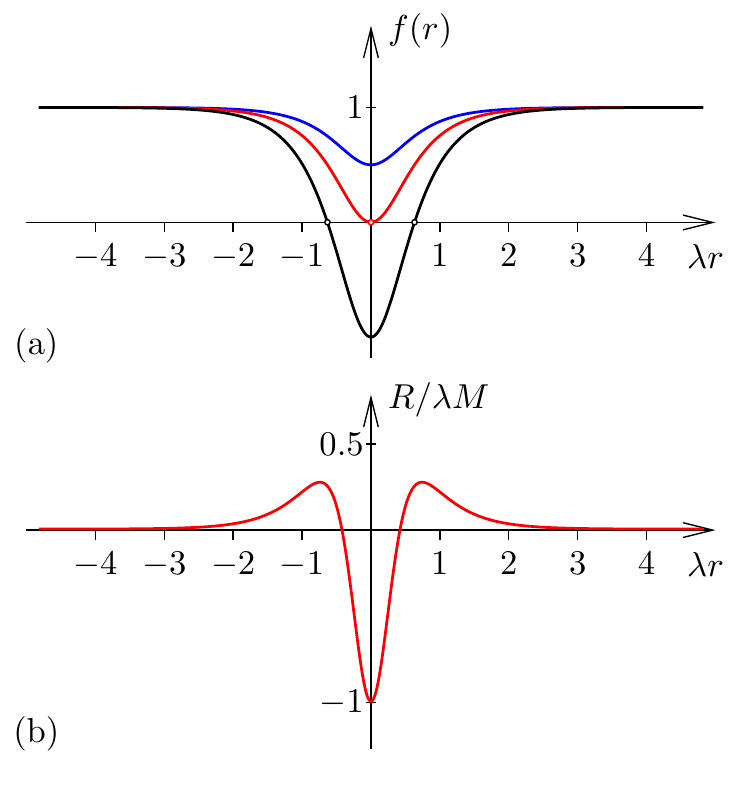}\hspace{0.5cm}\includegraphics[width=8.15cm]{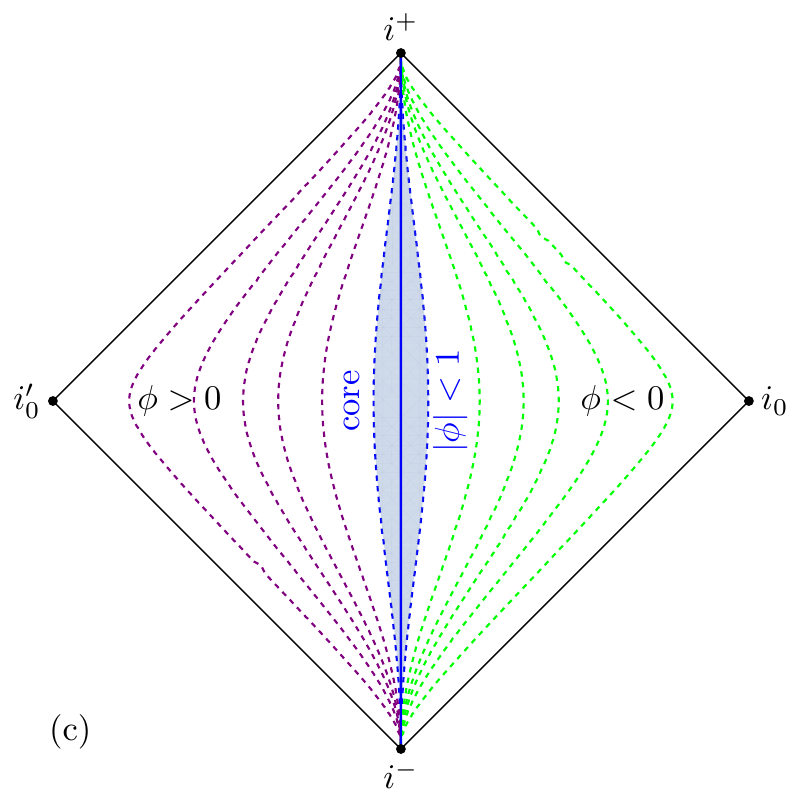}
}
%\vspace{-4mm} \hspace{3.5cm}(a) \hspace{7.5cm}(b)
\caption{ (a) Metric component $f(r)$ for different masses $M/\lambda=2,\,4,\,8$. (b) Ricci curvature. (c) Penrose diagram for the gravitational kink with $M<M_\mathrm{ext}=4\lambda$. } \label{fig:metric}
\end{figure}

The general vacuum solution \eqref{eq:general-et-bh} becomes one with
\begin{equation}\label{eq:sinh-f}
f(r)=1-\frac{M}{4\lambda\cosh(2\lambda r)}\;,
\end{equation}
The metric component~\eqref{eq:sinh-f} is plotted in Fig.~\ref{fig:metric}a for different values of the mass parameter $M$.

One finds the Ricci scalar scales uniformly with mass $M$. It concentrates inside the core reaching the maximum positive value $\sqrt{2/3^3}\cdot\lambda M$ at the borders and the minimum negative value $-\lambda M$ at the core's center, see in Fig.~\ref{fig:metric}b.

One notes from Eq.~\eqref{eq:sinh-f} there is a threshold mass $M_\mathrm{ext}=4\lambda$. If $M<M_\mathrm{ext}$ the metric~\eqref{eq:sinh-f} describes the gravitational kink depicted in Fig.~\ref{fig:metric}c. If $M>M_\mathrm{ext}$ the spacetime describes the non-extremal black bounce with outer and inner event horizons at $r=\pm r_{\mathrm{h}}$,
\begin{equation}
r_{\mathrm{h}}=\frac1{2\lambda}\arcosh\left(\frac{M}{M_\mathrm{ext}}\right)\;,
\end{equation}
so that $f(\pm r_{\mathrm{h}})=0$. For the extremal bounce one has $r_\mathrm{h}=0$.

Substituting $W(\phi)=-2\sinh(2\phi)$ into Eqs.~\eqref{gen-temp}, \eqref{gen-ent} one finds the black bounce entropy and temperature,
\begin{equation}
S_{BH}=\frac{2\pi}{\lambda}M\sqrt{1-\frac{M_\mathrm{ext}^2}{M^2}}\;.
\end{equation}
\begin{equation}\label{eq:sinh-temp}
T_H=\frac{\lambda}{2\pi}\sqrt{1-\frac{M_\mathrm{ext}^2}{M^2}}\;.
\end{equation}
For large black bounces the temperature approaches a value $\lambda/2\pi$ in agreement with the CGHS model limit.

The Schwarzschild coordinates $(t,\,r)$ in Eqs.~\eqref{eq:general-et-bh}, \eqref{eq:sinh-f} are geodesically incomplete and describe the exterior region at $r>r_{\mathrm{h}}$. In order to study the global structure of the eternal black bounce one needs to perform the coordinate extension. 

\begin{figure}[t]
\centerline{
\hspace{0.2cm}\includegraphics[width=7.87cm]{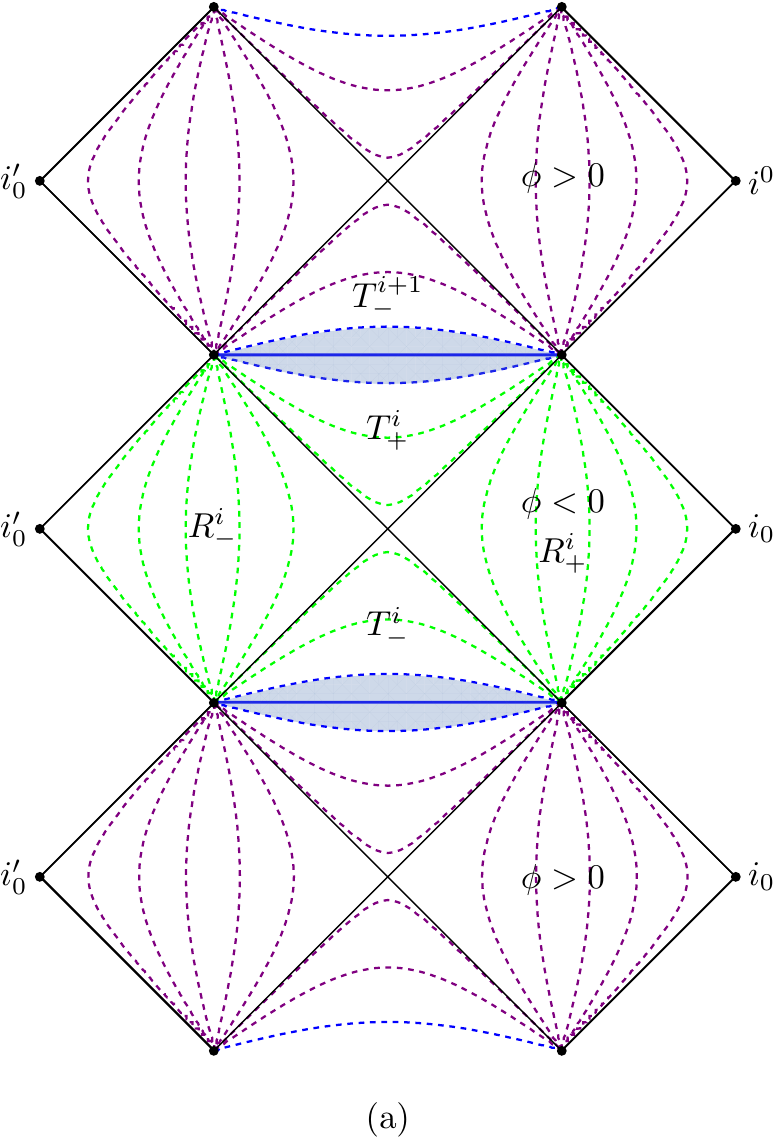}\hspace{0.5cm}\includegraphics[width=6.38cm]{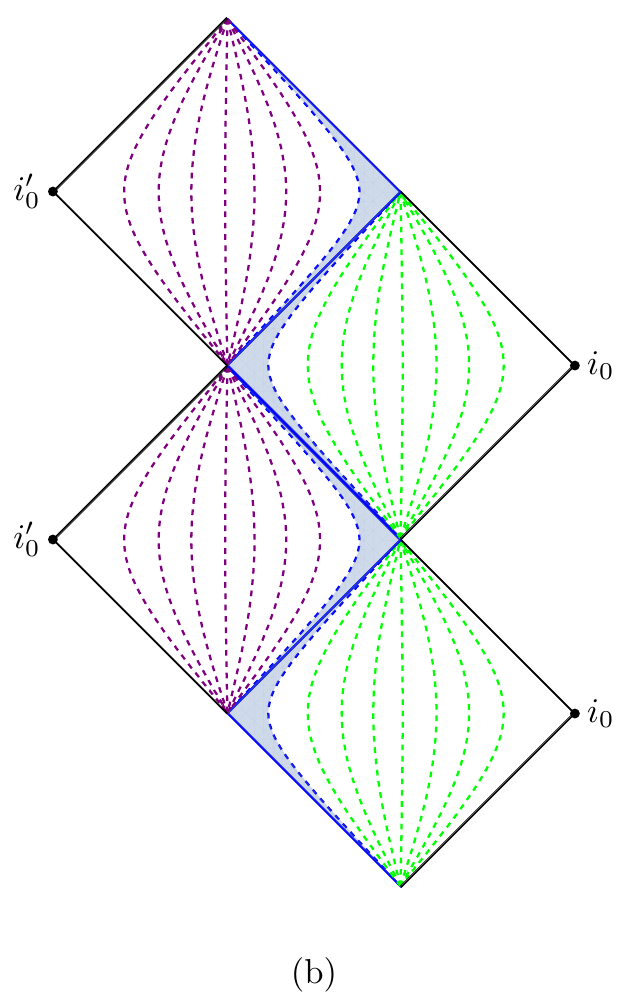}
}
%\vspace{-4mm} \hspace{3.5cm}(a) \hspace{7.5cm}(b)
\caption{ Penrose diagrams for (a) the eternal non-extremal bouce and (b) the extremal bounce (a similar picture arises for the model of Ref.~\cite{Ai:2020nyt}). } \label{fig:etbh-penrose}
\end{figure} 

Let us introduce new coordinates,
\begin{equation}\label{eq:coord-ext}
T=\sqrt{g(r)}\sinh(2\pi T_Ht)\;, \qquad R=\sqrt{g(r)}\cosh(2\pi T_Ht)\;,
\end{equation}
where
\begin{equation}
g(r)=\frac{\left(1+\frac{M_\mathrm{ext}}{M}\right)\tanh(\lambda r)-2\pi T_H/\lambda}{\left(1+\frac{M_\mathrm{ext}}{M}\right)\tanh(\lambda r)+2\pi T_H/\lambda}\exp(4\pi T_Hr)\;.
\end{equation}
The function $g(r)>0$ in the exterior region $r>r_{\mathrm{h}}$, so that the Schwarzschild coordinates $(t,\,r)$ cover a quadrant $R>|T|$. Applying the coordinate transformation~\eqref{eq:coord-ext} one finds the metric~\eqref{eq:general-et-bh}, \eqref{eq:sinh-f} in a form
\begin{equation}\label{eq:compl-metric}
ds^2=\frac{f(r)}{4\pi^2{T_H}^2g(r)}\left(-dT^2+dR^2\right)\;,
\end{equation}
where a conformal factor $\propto f(r)/g(r)$ is positive everywhere except at the inner horizon $r=-r_{\mathrm{h}}$.

The metric \eqref{eq:compl-metric} can be analytically continued into the interior region if one redefines coordinates $(T,\,R)$ as 
\begin{equation}
T=\sqrt{-g(r)}\cosh(2\pi T_Ht)\;, \qquad R=\sqrt{-g(r)}\sinh(2\pi T_Ht)\;,
\end{equation}
so that $(T,\,R)$ cover the entire spacetime patch with $r>-r_\mathrm{h}$.

This patch is still not geodesically complete. The global spacetime consists of the infinite number of patches with the interior regions matched onto each other. The lines $\phi=\mathrm{const}$ belonging to $T_+$-region of i-th patch and $T_-$-region of (i+1)-th patch can be identified by a map
\begin{equation}\label{eq:map}
V_{i+1}=-\frac{\kappa}{V_i}\;, \qquad\qquad U_{i+1}=-\frac1{\kappa U_i}\;,
\end{equation}
where $V_i=T_i+R_i$, $U_i=T_i-R_i$ are the light-cone coordinates on the i-th patch, and $\kappa$ is a residual parameter corresponding to respective shifts of the identified patches along the lines $\phi=\mathrm{const}$. The resulting Penrose diagram is presented in Fig.~\ref{fig:etbh-penrose}. Unlike traversable wormhole observers can travel in one time direction only\footnote{An alternative term timehole was introduced to describe such spacetimes in Ref.~\cite{Visser:1997yn}.}.

%%%%%%%%%%%%%%%%%%%%%%%%%%%%%%%%
\section{Matter considerations}\label{sec:matter-consid}
\subsection{Core stability?}
One may wonder if a singularity appears in the core region ($|\phi|\lesssim 1$) in response to a perturbation by infalling matter. We consider one of the massless scalar fields $f_i\equiv f$ in Eq.~\eqref{eq:cghs-action-n} as a source of gravity with the stress tensor,
\begin{equation}
{T_\mathrm{m}}_{\mu\nu}=\nabla_\mu f\nabla_\nu f-\frac12g_{\mu\nu}(\nabla f)^2\;,
\end{equation}
on the r.h.s of Eq.~\eqref{eq:ldv2}. Notice that the matter appears to anti-gravitate from the viewpoint of asymptotic observer on the left side ($\phi>0$) of the spacetime because of change of sign in $W(\phi)$. Nevertheless, it does not make necessarily the model unstable because matter fields have manifestly positive energy and gravity sector does not have propagating degrees of freedom in two dimensions.

\begin{figure}[t]
\centerline{
\includegraphics[width=6.03cm]{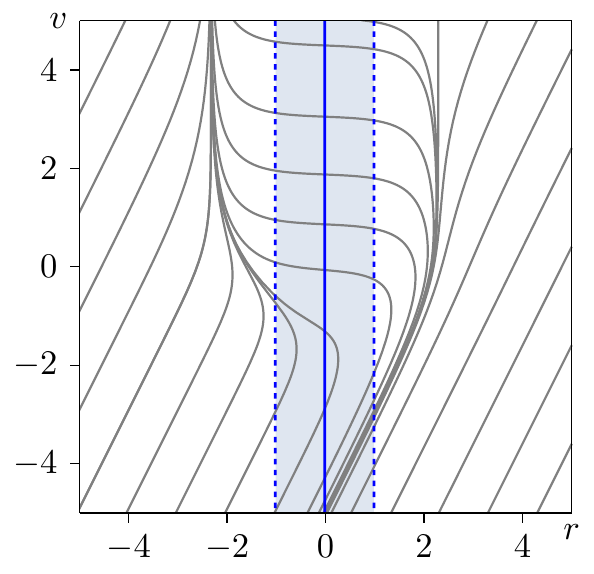}
}
\caption{ The Vaidya solution~\eqref{eq:vaidya} with a wave packet $\partial_vf(v)=10\lambda\,\mathrm{sech}(\lambda v)$ (blurry horizontal strip). Dark lines are outgoing null geodesics. The inner and outer horizons are correspondingly a future and past attractors for null geodesics. } \label{fig:vaidya}
\end{figure}

Ingoing wave packet $f(v)$ admits an exact solution with the Vaidya metric,
\begin{eqnarray}\label{eq:vaidya}
ds^2=-F(v,r)dv^2+2dvdr\;, \notag\\
F(v,r)=1-\frac{{\cal M}(v)}{4\lambda\cosh(2\lambda r)}\;, \qquad \partial_v{\cal M}(v)=(\partial_vf(v))^2\;,
\end{eqnarray}
where ${\cal M}(v)$ is the Bondi mass, see Fig.~\ref{fig:vaidya}. One finds the Ricci scalar $R=-\partial_r^2F(v,r)$ is everywhere finite and concludes the core is classically stable.

We need to remark that classical stability of the core does not mean automatically the singularity is absent in the full quantum regime. Indeed, the core region lives formally at the strong coupling as $W(\phi)\to 0$ and the quantum corrections can drastically modify its internal structure.

One may wonder if something peculiar happens already after inclusion of the one-loop corrections from the matter fields \cite{Polyakov:1981rd}. It was argued that the classical linear dilaton vacuum is unstable in the CGHS model. The semiclassical static solutions were found with the dilaton field bouncing off the strong coupling region \cite{Birnir:1992by}.

It is intriguing if the similar thing could happen in the models with black bounces so that the strong coupling problem is avoided. Unfortunately, we found this is not the case, see Appendix~\ref{app:kinks} for details. In this paper we assume that singularity never appears and the strong coupling poses no threat.

A minisuperspace approximation of the RST model suggests there is a bouncing behaviour \cite{Daghigh:2018qte}. The similar picture is motivated by loop quantum gravity (LQG)~\cite{Ashtekar:2020ifw}. Nevertheless, there can be severe obstacles to adopt such a picture without hesitation, namely LQG provides generic mechanisms which may turn the singularity into an acausal ``euclidean core'' free of infinite curvature but impeding deterministic evolution \cite{Bojowald:2014zla}. 

\subsection{Mass inflation?}
There is a widespread opinion that the regular black holes with Cauchy horizons are unstable because of the mass inflation phenomenon \cite{Poisson:1989zz}. This effect is related to exponential accumulation of matter near inner horizons leading to formation of the spacelike singularity. 

This instability may not be necessarily present \cite{Dokuchaev:2013uda,Bonanno:2020fgp} but it was demonstrated that charged black holes in the CGHS model suffer from the mass inflation both at classical and semiclassical levels \cite{Balbinot:1994ee,Chan:1994tb,Frolov:2006is}. Therefore, it becomes important to check whether the inner Cauchy horizons are stable against matter perturbations in our case.

We assume that black bounces are large $M\gg M_\mathrm{ext}$ and ignore evaporation effects. We approximate spacetime with the CGHS model solution in the regions far from the core at $|\phi|\lesssim 1$. The general CGHS solution with choice of the metric $ds^2=-e^{2\rho}dvdu$ is
\begin{eqnarray}
e^{-2\rho}=e^{-2\phi}=-\lambda^2 vu+g(v)+h(u)\;, \notag\\
g(v)=\frac12\int_0^vdv'\int^{+\infty}_{v'}dv''(\partial_vf(v''))^2\;,  \label{eq:cghs-exact}\\
h(u)=-\frac12\int_{-\infty}^udu'\int_{-\infty}^{u'}du''(\partial_uf(u''))^2\;, \notag
\end{eqnarray}
where $g(v),\,h(u)$ are functions depending on the matter content \cite{Callan:1992rs}.

One takes
\begin{equation}
g(v)\simeq\frac{M}{2\lambda}-\frac{g_\infty}{(\lambda v)^{2\alpha}}\;, \qquad\qquad \alpha>0\;,
\end{equation}
which corresponds to a power--law tail $f(v)\simeq f_0\cdot(\lambda v)^{-\alpha}$ as $v\to +\infty$. After passing the core region the wave packet distorts accordingly to Eq.~\eqref{eq:map},
\begin{equation}\label{eq:ftof}
f(v)\,\mapsto\,f_0\cdot(-\lambda v)^\alpha\;,
\end{equation}
on the future side of the black bounce.

Substituting~\eqref{eq:ftof} into Eq.~\eqref{eq:cghs-exact} and using $R=-2\Box\rho$ one obtains the Ricci scalar,
\begin{multline}
R\simeq 4\lambda^2 e^{2\phi}\left(\frac{M}{2\lambda}+(2\alpha+1)g_\infty(-\lambda v)^{2\alpha}+\frac{{\cal E}_{\mathrm{out}}(u)}{2\lambda}\right. \\
\left.+\frac{2\alpha+1}{2\alpha-1}\frac{2\alpha g_\infty}{\lambda}(-\lambda v)^{2\alpha-1}\partial_uh(u)\right)\;, \label{eq:curva-mass-inflation}
\end{multline}
to be finite at the Cauchy horizon if $\alpha>1/2$. Otherwise, the outgoing wave packet crossing the Cauchy horizon at $v=0$ triggers a singularity formation.

Concerning regularity of the solution across the Cauchy horizon Eq.~\eqref{eq:curva-mass-inflation} does not apply at $v>0$ because a continued solution depends on the undetermined matter content in the future spacetime patch. The whole solution is not analytic because of arbitrariness in choice of the wave packets living on the future side of the bounce. One can only assume that the solution~\eqref{eq:cghs-exact} describes the black hole with fixed mass at $v\to +0$, so that 
\begin{equation}
g(v)\simeq\frac{M}{2\lambda}+g_0\cdot(\lambda v)^{2\alpha}\;, \qquad\qquad \alpha>0\;.
\end{equation}
It is only possible if the wave packets propagating along the Cauchy horizon at $v\to +0$ have finite energy and there is no singularity. More importantly all the problems associated with the mass inflation may be resolved by backreaction from matter fields on geometry as we argue in the next Section.

\subsection{Remnants?}
\label{sec:remnants}
In the adiabatic approximation the Hawking radiation from black bounces is approximately thermal described by a black body spectrum. The radiation flux can be related to a mass change rate by the 2D Stefan–Boltzmann law, see Appendix~\ref{app:conf-anom} for derivation,
\begin{equation}\label{eq:balance}
\frac{dM}{dt}=-\frac{\pi}{12}T_H^2(M)\;,
\end{equation}
where temperature $T_H(M)$ is given by Eq.~\eqref{eq:sinh-temp}.

\begin{figure}[t]
\centerline{
\hspace{0.25cm}\includegraphics[width=7.7cm]{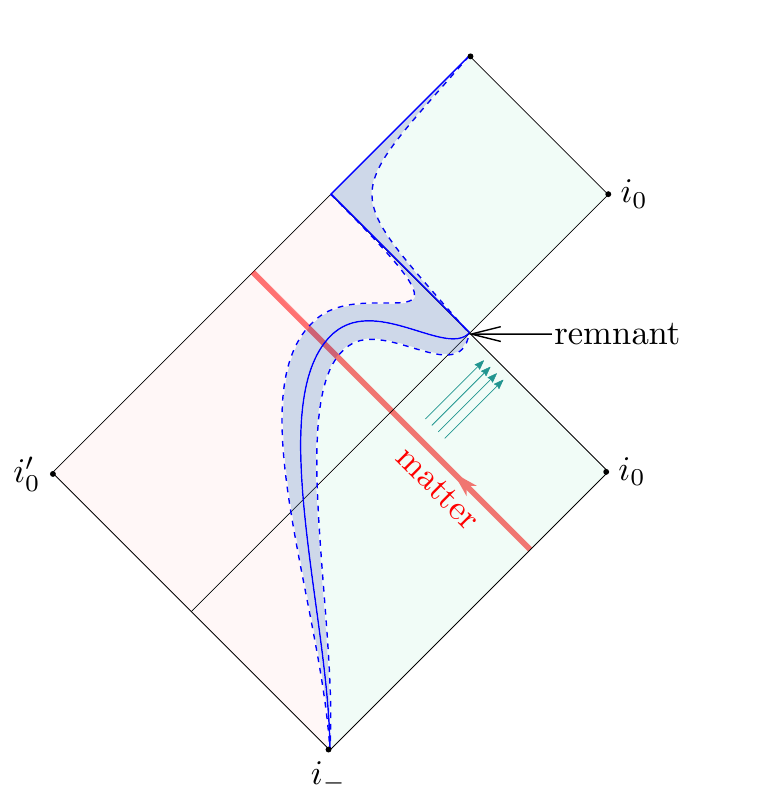}\hspace{-0.5cm}\includegraphics[width=8.14cm]{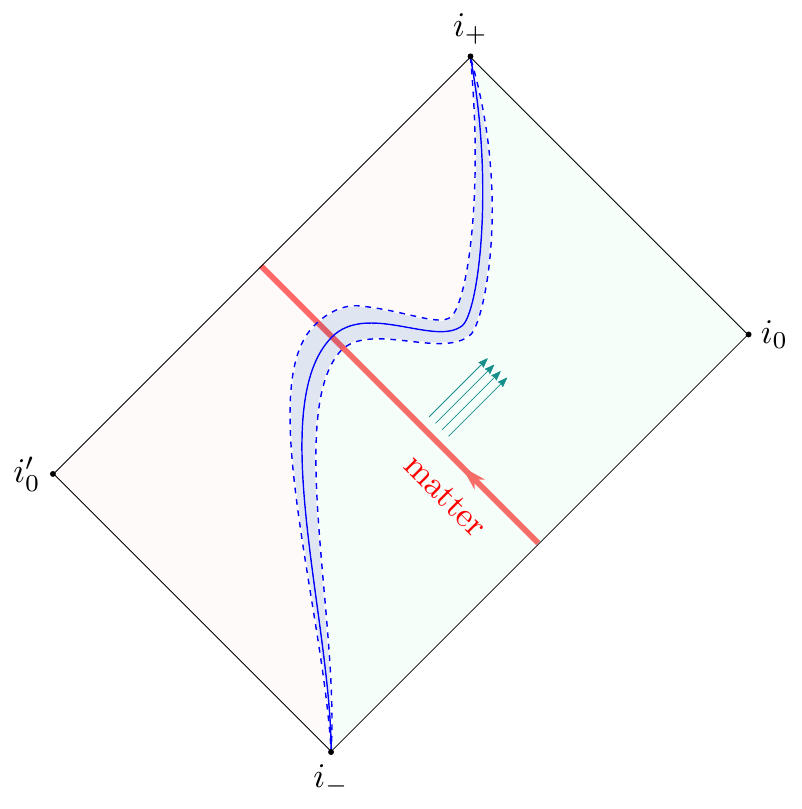}
}
\vspace{0mm} \hspace{2.75cm}(a) \hspace{5.5cm}(b)
\caption{ Penrose diagrams for (a) the evaporating black bounce formed from gravitational collapse of matter in the semiclassical approximation and (b) the proposed scenario for decaying remnant in the end of evaporation. Arrows represent Hawking radiation. } \label{fig:remnant-diagram}
\end{figure}

One integrates out Eq.~\eqref{eq:balance} and obtains
\begin{equation}\label{eq:evap-process}
M(t)+\frac{M_\mathrm{ext}}{2}\log\left(\frac{M(t)-M_\mathrm{ext}}{M(t)+M_\mathrm{ext}}\right)=M_0-\frac{\lambda^2t}{48\pi}\;,
\end{equation}
where $M_0\gg M_\mathrm{ext}$ was assumed. Early times are characterised by the linear regime of evaporation corresponding to the CGHS model according to Eq.~\eqref{eq:evap-process}. After loosing the bulk of its mass evaporation of the black bounce slows down,
\begin{equation}\label{eq:remnant-forms}
M\simeq M_\mathrm{ext}\left(1+\exp\left(-\frac{\lambda^2 t}{24\pi M_\mathrm{ext}}\right)\right)\;,
\end{equation}
so it takes infinite amount of time to settle down to the extremal limit. This configuration seems to be stable and can be regarded as a remnant, see in Fig.~\ref{fig:remnant-diagram}a. Possibility for remnants in the CGHS model was previously advocated in Ref.~\cite{Almheiri:2013wka}.

One suspects it is unreliable to draw conclusion about stability of the remnants from the semiclassical picture because a precise dynamics at the late times \eqref{eq:remnant-forms} should depend on the strong coupling behaviour.

Indeed, one conjectures the thermal fluctuations of the fields near the event horizon can destroy the remnant. A characteristic decay time can be estimated from thermodynamic principles as follows. Recalling Einstein theory of fluctuations,
\begin{equation}
\langle M\rangle=-\frac1{Z}\frac{\partial Z}{\partial\beta}\;, \qquad \langle M^2\rangle=\frac1{Z}\frac{\partial^2Z}{\partial\beta^2}\;, \qquad\Rightarrow\quad \langle(\Delta M)^2\rangle=-\frac{\partial\langle E\rangle}{\partial\beta}\;,
\end{equation}
one finds using Eq.~\eqref{eq:sinh-temp} a deviation
\begin{equation}
\delta M=M_\mathrm{dec}-M_\mathrm{ext}=\sqrt{\langle(\Delta M)^2\rangle}=\frac{\lambda^2}{M_\mathrm{ext}}{\displaystyle O}(1)
\end{equation}
from the mean field result \eqref{eq:remnant-forms} assuming $\delta M\ll M_\mathrm{ext}$. Substitution into Eq.~\eqref{eq:remnant-forms} gives an estimate, 
\begin{equation}
t_\mathrm{dec}\simeq 48\pi\frac{M_\mathrm{ext}}{\lambda^2}\log\left(\frac{M_\mathrm{ext}}{\lambda}\right)\;,
\end{equation}
for the expected remnant lifetime.

Hereby we propose a scenario for the entire evolution. The black bounce approaching the extremal state is subjected to stochastic quantum field fluctuations which cause transition from collapsing into expanding near-extremal black bounce. This state releases the would-be interior matter content into the same region of spacetime rendering unitary evolution and then relaxes to a gravitational kink state of unspecified mass, see Fig.~\ref{fig:remnant-diagram}b. The aforementioned mass inflation problem also becomes removed because the Cauchy horizon disappears too in this picture.

%%%%%%%%%%%%%%%%%%%%%%%%%%%%%%%%%
\section{Discussion}\label{sec:discussion}
We propose simple two-dimensional dilaton gravity models with non-singular black holes and Minkowski spacetime as a trivial solution. The properties of non-trivial vacuum solution depends on its mass. There are gravitational kinks ($M<M_\mathrm{ext}$), extremal ($M=M_\mathrm{ext}$) and non-extremal ($M>M_\mathrm{ext}$) black bounces. The gravitational kinks has the same causal structure as the Minkowski spacetime. The eternal bounces resemble charged black holes with separate asymptotically flat regions connected by one-way interiors.

We considered a particular model approximated by the CGHS model except for the central core region of size $\lambda^{-1}$. This core stable against perturbations by classical infalling matter. We found that mass inflation on the Cauchy horizons can happen at certain choice of initial conditions.

Euclidean methods allows to calculate temperature and entropy of the bounces. The extremal black bounces are supposed to have zero temperature and entropy hence they can be considered as remnants.

We argue that remnants are unstable because of the thermal fluctuations of mass $\delta M\simeq T_H$. Infinitesimally small change can completely dissolve the event horizon. This scenario of evaporation is analogous to one proposed for regular black holes in Refs.~\cite{Hayward:2005gi,Frolov:2014jva}. Loop quantum gravity motivates the similar picture of evaporation for real black holes \cite{Bianchi:2018mml}.

Let us discuss some prospects for future research. Next step will be further development of the S-matrix approach, e.g. in the spirit of 't Hooft's black hole ansatz \cite{Stephens:1993an}. 

Previously, we calculated semiclassical scattering amplitudes using regularization method in the CGHS model with a reflecting boundary $\phi=\phi_0$ and a massive pointlike particle as matter \cite{Fitkevich:2020tcj}, see also Refs.~\cite{Fitkevich:2017izc,Fitkevich:2020okl} for more details on this model. Obtained results are consistent with unitarity but analogous calculation for complete theory with quantum matter fields is still missing.

The same method can be applied to the sinh-CGHS model in attempt to find the regular solution contributing to the path integral. One may expect that given initial and final quantum states, namely high-energy collapsing wave packet and outgoing low-energy Hawking radiation, regularization method produces a horizonless spacetime envisioned in Sec.~\ref{sec:remnants}.

Another interesting path for research is to consider AdS$_2$/CFT$_1$ correspondence in the spirit of the AP model \cite{Almheiri:2014cka}. Addition of the term
\begin{equation}
S_{\mathrm{AdS}_2}=-8\Lambda\int d^2x\sqrt{-g}\left(\phi\cosh(2\phi)+\phi^2\sinh(2\phi)\right)
\end{equation}
to the action \eqref{eq:grav-action} replaces vacuum solution \eqref{eq:sinh-f} with $f(r)=1-M/4\lambda\cosh(2\lambda r)+\Lambda r^2$ so that it describes asymptotically AdS$_2$ background. Understanding of gravitational scattering in terms of the dual boundary theory may shed more light on holography in realistic models of quantum gravity.

Additional important question concerns the ``islands'' method for calculating the entanglement entropy \cite{Almheiri:2019qdq}. It seems reasonable to connect the replica wormholes with the saddle-point solutions contributing to scattering amplitudes. Models with non-singular black holes are perfectly suited for this task.

%%%%%%%%%%%%%%%%%%%%%%%%%%%%%%%%
\paragraph{Acknowledgments.} We thank Dmitry Levkov, Andrey Shkerin, Valery Rubakov, and Andrei Kataev for inspiring discussions. This work was supported by the grant 21-1-4-11-1 of the Foundation for the Advancement of Theoretical Physics and Mathematics ``BASIS''. 

%%%%%%%%%%%%%%%%%%%%%%%%%%%%%%%%
\appendix
%%%%%%%%%%%%%%%%%%%%%%%%%%%%%%%% 
\section{Quantum kinks in sinh-CGHS}
\label{app:kinks}
In this Appendix we investigate static solutions in the model~\eqref{eq:grav-action} with the one-loop corrections from $N$ matter fields. It is accounted by adding the
Polyakov action \cite{Polyakov:1981rd}, which can be recast in the local form,
\begin{equation}\label{eq:polyakov}
S_\mathrm{one-loop}=\int d^2x\sqrt{-g}\left(-\frac12(\nabla\chi)^2+\sqrt{\frac{N}{48\pi}}\chi R\right)\;,
\end{equation}
by using an auxiliary field $\chi$ representing contribution from all $N$ scalar fields given vacuum background state. By varying Eq.~\eqref{eq:polyakov} one finds the field equation,
\begin{equation}
\Box\chi=-\sqrt{\frac{N}{48\pi}}R\;,
\end{equation}
and the energy-momentum tensor,
\begin{equation}
T^\chi_{\mu\nu}=\nabla_\mu\chi\nabla_\nu\chi-\frac12g_{\mu\nu}(\nabla\chi)^2+\sqrt{\frac{N}{12\pi}}\left(\nabla_\mu\nabla_\nu\chi-g_{\mu\nu}\Box\chi\right)\;,
\end{equation}
on the r.h.s of Eq.~\eqref{eq:ldv2}.

With choice of the metric with the line element $ds^2=e^{2\alpha(r)}(-dt^2+dr^2)$ one derives differential equations,
\begin{eqnarray}
W'(\phi(r))\alpha''(r)+W''(\phi(r))\phi''(r)+\frac12 W'''(\phi(r))\left((\phi'(r))^2-\lambda^2 e^{2\alpha(r)}\right)=0\;, \label{eq:num1}\\
\frac{N}{24\pi}\alpha''(r)+W'(\phi(r))\phi''(r)+W''(\phi(r))\left((\phi'(r))^2-\lambda^2 e^{2\alpha(r)}\right)=0\;, \label{eq:num2}
\end{eqnarray}
for unknown functions $\phi(r)$, $\alpha(r)$. Initial conditions are fixed by asymptotic behaviour of fields.

\begin{figure}[t]
\centerline{
\hspace{0.2cm}\includegraphics[width=6.38cm]{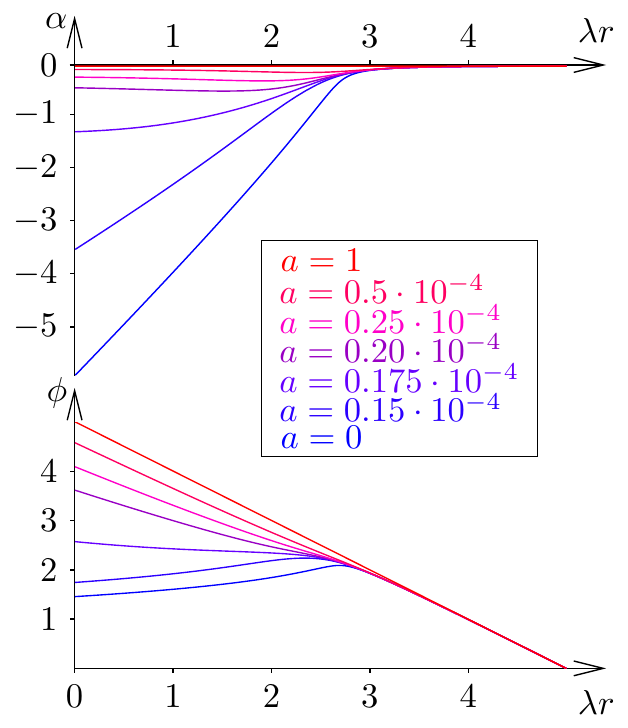}\hspace{0.5cm}\includegraphics[width=6.38cm]{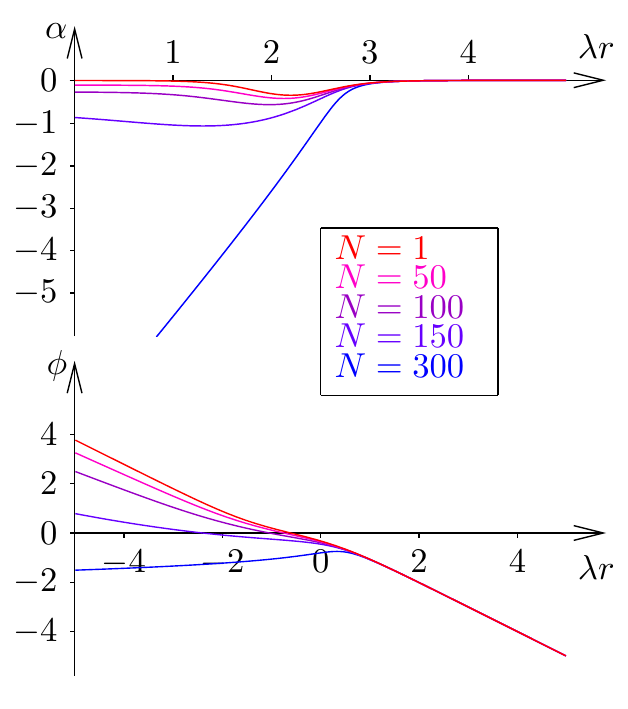}
}
\vspace{-0mm} \hspace{4.5cm}(a) \hspace{6.0cm}(b)
\caption{ (a) Set of numerical solutions interpolating between the CGHS and sinh-CGHS models. Mass $M=10^{-3}\cdot M_\mathrm{ext}$ as seen by asymptotic observer and number of fields $N=1$. (b) Transition from classical kink of mass $M=0.5\cdot M_\mathrm{ext}$ to quantum kink in the sinh-CGHS model with growing number of quantum fields $N$. } \label{fig:qkink}
\end{figure}

Let us consider Eq.~\eqref{eq:gen-lin-dil} with $W(\phi)=e^{-2\phi}-a\cdot e^{2\phi}$ interpolating between the CGHS ($a=0$) and sinh-CGHS ($a=1$) models. We solved Eqs.~\eqref{eq:num1}, \eqref{eq:num2} numerically on {\it Mathematica} using the implicit Runge-Kutta method. We present results in Fig.~\ref{fig:qkink}.

Numerical solution with $a=0$ reproduces quantum kinks obtained in Ref.~\cite{Birnir:1992by}. Analogous picture arises in the sinh-CGHS models if the number of scalar fields is comparably large. In both cases the Ricci scalar $R=-2e^{-2\alpha}\partial_r^2\alpha$ diverges as $r\to -\infty$ and the quantum kink has singularity at the left infinity. Regular solutions do not have bouncing behaviour of dilaton field.

\section{Hawking radiation from Weyl anomaly}
\label{app:conf-anom}
We revisit here calculation of the renormalized energy-momentum tensor of the massless scalar field in two dimensions. One recalls the expectation value of its trace is anomalous,
\begin{equation}
\langle\hat{T}^\mu_\mu\rangle_\psi=-\frac{R}{24\pi}\;,
\end{equation}
given a quantum state of the matter. 

For the metric with a line element $ds^2=-f(r)dvdu$ where $r$ is an implicit function of $v-u$ one finds
\begin{gather}
\langle T_{vv}\rangle_\psi=\frac1{96\pi}\left(f''(r)f(r)-\frac12\left(f'(r)\right)^2\right)+g_\psi(v)\;, \\
\langle T_{uu}\rangle_\psi=\frac1{96\pi}\left(f''(r)f(r)-\frac12\left(f'(r)\right)^2\right)+h_\psi(u)\;,
\end{gather}
where $g_\psi(v)$ and $h_\psi(u)$ are function determined by the quantum state. 

The Unruh state is given by
\begin{equation}
g_\mathrm{Unruh}(v)=0\;, \qquad h_\mathrm{Unruh}(u)=\frac{\lambda^2}{48\pi}\left(1-\frac{M_\mathrm{ext}^2}{M^2}\right)\;,
\end{equation}
where we used Eq.~\eqref{eq:sinh-f}. This corresponds to the Hawking flux from the collapsing black bounce with $M>M_\mathrm{ext}$. Calculation shows $\langle T_{uu}\rangle_\mathrm{Unruh}$ is regular at the event horizon and future infinity where it gives a black body radiation flux with the temperature~\eqref{eq:sinh-temp}.

%One may wonder what happens in transition between gravitational kinks with different masses.
%\begin{equation}
%\label{reg-et-bh}
%ds^2=\left(1-\frac{M}{4\lambda\cosh(2\lambda r)}\right)\left(-dt^2+d\tilde{r}^2\right)\;,
%\end{equation}
%where we introduced a ``tortoise'' coordinate
%\begin{equation}
%\tilde{r}=r+\frac1{\lambda\sqrt{\left(\frac{4\lambda}{M}\right)^2-1}}\arctan\left(\frac{1+M/4\lambda}{\sqrt{1-(M/4\lambda)^2}}\tanh(\lambda r)\right)\;.
%\end{equation}
%Coordinate systems $(t,\,r)$ and $(t,\,\tilde{r})$ both cover exterior region $r>r_{\mathrm{h}}$ only and incomplete.
%%%%%%%%%%%%%%%%%%%%%%%%%%%%%%%%%%%%%%%%%

\end{document}